\documentclass[showpacs]{revtex4}
\usepackage{graphicx,psfrag,amsmath,amssymb,amsfonts,bbm,latexsym,color,dcolumn}
\usepackage{graphicx,psfrag,amsmath,amssymb,amsfonts,bbm,latexsym,color,dcolumn}

\newcommand{\be}{\begin{equation}}
\newcommand{\bea}{\begin{eqnarray}}
\newcommand{\ee}{\end{equation}}
\newcommand{\eea}{\end{eqnarray}}

\input{tcilatex}

\begin{document}

\date{November 2008}
\title{Asymptotically AdS$_3$ Solutions to Topologically Massive Gravity at
Special Values of the Coupling Constants}
\author{Alan Garbarz, Gaston Giribet}
\affiliation{Departamento de F\'{\i}sica, Universidad de Buenos Aires, Ciudad
Universitaria, Pabell\'{o}n I, 1428 Buenos Aires, Argentina.}
\author{Yerko V\'{a}squez}
\affiliation{Instituto de F\'{\i}sica, Pontificia Universidad Cat\'{o}lica de Valpara%
\'{\i}so, Casilla 4950, Valpara\'{\i}so, Chile.}

\begin{abstract}
We study exact solutions to Cosmological Topologically Massive Gravity
(CTMG) coupled to Topologically Massive Electrodynamics (TME) at special
values of the coupling constants. For the particular case of the so called
chiral point $l\mu _G=1$, vacuum solutions (with vanishing gauge field) are
exhibited. These correspond to a one-parameter deformation of GR solutions,
and are continuously connected to the extremal Ba\~{n}%
ados-Teitelboim-Zanelli black hole (BTZ) with bare constants $J=-lM$. At the
chiral point this extremal BTZ turns out to be massless, and thus it can be
regarded as a kind of \textit{ground state}. Although the solution is not
asymptotically AdS$_3$ in the sense of Brown-Henneaux boundary conditions,
it does obey the weakened asymptotic recently proposed by Grumiller and
Johansson. Consequently, we discuss the holographic computation of the
conserved charges in terms of the stress-tensor in the boundary. For the
case where the coupling constants satisfy the relation $l\mu_G=1+2l\mu_E$,
electrically charged analogues to these solutions exist. These solutions are
asymptotically AdS$_3 $ in the strongest sense, and correspond to a
logarithmic branch of selfdual solutions previously discussed in the
literature.
\end{abstract}

\maketitle







\section{Introduction}

In the last year and a half there has been a revived interest in
three-dimensional gravity. This was mainly motivated by E. Witten's proposal 
\cite{witten} that Einstein gravity in AdS$_{3}$ is holographically dual to
a holomorphically factorizable CFT$_{2}$. This idea has attracted
considerable attention, and led to intense debate \cite%
{witten2,gaberdiel,gaiotto,gaberdiel2}. Another model of three-dimensional
gravity that has attracted much attention recently is Topologically Massive
Gravity (TMG), which corresponds to three-dimensional Einstein gravity
coupled to a gravitational Chern-Simons term without torsion \cite{DJT};
namely, 
\begin{equation}
I_{G}=\frac{1}{2\kappa ^{2}}\int_{\mathcal{M}}d^{3}x\sqrt{-g}(R+\frac{2}{%
l^{2}})+\frac{1}{4\kappa ^{2}\mu _{G}}\int_{\mathcal{M}}d^{3}x\epsilon
^{\lambda \mu \nu }\Gamma _{\lambda \sigma }^{\rho }(\partial _{\mu }\Gamma
_{\rho \nu }^{\sigma }+\frac{2}{3}\Gamma _{\mu \rho }^{\gamma }\Gamma _{\nu
\gamma }^{\sigma })+\mathcal{B}  \label{Action}
\end{equation}%
with $l^{-2}=-\Lambda $ and $\kappa ^{2}=8\pi G$, and where $\mathcal{B}$
stands for the boundary term which we are not writing explicitly here (see (%
\ref{ups}) below). The three-dimensional gravity theory defined by (\ref%
{Action}) contains a local massive graviton degree of freedom \cite{DJT,DJT2}%
, and it also admits black hole solutions \cite{MCL}, what makes TMG a very
interesting model to be explored.

One of the interesting properties of TMG is that its holographic description 
\cite{KL} in terms of a CFT$_{2}$ captures several interesting features of
the AdS$_{3}$/CFT$_{2}$ realization. As in the case of Einstein gravity in
AdS$_{3}$, the asymptotic isometry group of TMG in this background is
generated by two copies of the Virasoro algebra with non-trivial central
extension. When the gravitational Chern-Simons coupling $\mu _{G}$ takes the
special value $\mu _{G}=1/l$, the central charge of left-moving excitations
in the boundary theory vanishes, leading to the still controversial
suggestion that the theory might be chiral \cite{LSS2}; see also\ \cite{CDWW}%
-\cite{Desernuevo}.

In addition to AdS$_{3}$, other backgrounds of TMG have recently shown to be
of great interest. In particular, warped versions of AdS$_{3}$ have led to
fabulous applications such as the description of extremal four-dimensional
Kerr black holes \cite{Suecos,GHSS,ALPSS,CD,Pope,Stromingernuevo}. The
connection of these backgrounds to G\"{o}del black holes \cite{CD}\ are also
very interesting.

Here, we will be concerned with Topologically Massive Gravity coupled to its
electromagnetic analog, the Topologically Massive Electrodynamics (TME). The
gauge theory action is given by the Maxwell term coupled to abelian
Chern-Simons term; namely%
\begin{equation}
I_{E}=-\frac{1}{4}\int_{\mathcal{M}}d^{3}x\sqrt{-g}F_{\mu \nu }F^{\mu \nu }+%
\frac{\mu _{E}}{4}\int_{\mathcal{M}}d^{3}x\epsilon ^{\mu \nu \rho }A_{\mu
}F_{\nu \rho }.  \label{A2}
\end{equation}

In this paper, we will consider the special case $l\mu _{G}=1-2\epsilon $
with $\epsilon =-l\mu _{E}$. The reason why we are particularly interested
in this relation between coupling constants is that such theories admit a
class of solution with interesting properties. For instance, particular
features of exact solutions at $l\mu _{G}=1$ ($\varepsilon =0$) were noticed
even before the \textit{chiral gravity conjecture} \cite{LSS} was
formulated; see for instance \cite{eloy}. At these points of the space of
parameters several exact solutions reported in the literature are seen to
coincide, and it is precisely when this degeneracy happens that new
solutions with interesting properties usually come up. In particular, at the
chiral point the solutions we will describe here correspond to an
asymptotically AdS$_{3}$ solutions of TMG in vacuum (with vanishing gauge
field). We discuss the theory at the chiral point in Section II, where we
present these vacuum solutions and discuss their properties in detail. In
Section III, we generalize the solutions to the case of TMG charged under
TME theory with $l\mu _{G}=1+2l\mu _{E}$. The charged solutions turn out to
be asymptotically AdS$_{3}$, with a gauge field configuration that diverges
at finite radius. We also discuss the relation between the solutions we
present here with self-dual solutions previously reported in the literature.
We summarize the results in Section IV.

\section{Vacuum solutions at the chiral point}

\subsection{Topologically massive gravity and its solution}

The equations of motion of TMG follow from varying (\ref{Action}) with
respect to the metric $g_{\mu \nu }$. In presence of matter (consider in
particular (\ref{A2})), these equations read%
\begin{equation}
R_{\mu \nu }-\frac{1}{2}Rg_{\mu \nu }-\frac{1}{l^{2}}g_{\mu \nu }+\frac{1}{%
\mu _{G}}C_{\mu \nu }=\kappa ^{2}T_{\mu \nu },  \label{caritaz}
\end{equation}%
where $\Lambda =-l^{2}$, $T_{\mu \nu }$ is the stress-tensor of the
electromagnetic field, and $C_{\mu \nu }$ is the Cotton tensor, given by%
\begin{equation}
C_{\mu \nu }=\frac{1}{2}\varepsilon _{\mu }^{\ \alpha \beta }\nabla _{\alpha
}R_{\beta \nu }+\frac{1}{2}\varepsilon _{\nu }^{\ \alpha \beta }\nabla
_{\alpha }R_{\mu \beta }.  \label{Cotono}
\end{equation}

In three dimensions the Weyl tensor identically vanishes, and the Cotton
tensor is the one that comes to play its role: It is a traceless tensor that
vanishes if and only if the metric is locally conformally flat. Traceless
condition implies that all the solutions of the field equations satisfy 
\begin{equation*}
R=-\frac{6}{l^{2}}-2\kappa ^{2}T_{\mu }^{\ \mu },
\end{equation*}%
and one finds that all three-dimensional Einstein manifolds solve (\ref%
{caritaz}).

Let us begin by considering the theory at the chiral point $l\mu _{G}=\pm 1$%
. At this point, we can consider solutions with vanishing gauge field, and
the coupling $\mu _{E}$ then takes an arbitrary value. More precisely, at
the chiral point $l\mu _{G}=1$ one finds a vacuum solution of TMG, whose
metric reads%
\begin{equation}
ds^{2}=-N^{2}(r)dt^{2}+\frac{dr^{2}}{N^{2}(r)}+r^{2}(N_{\phi }(r)dt-d\phi
)^{2}+N_{k}^{2}(r)(dt-ld\phi )^{2}  \label{uncharged}
\end{equation}%
where%
\begin{equation}
N^{2}(r)=\frac{r^{2}}{l^{2}}-\kappa ^{2}M+\frac{\kappa ^{4}M^{2}l^{2}}{4r^{2}%
},\qquad \qquad N_{\phi }(r)=\frac{\kappa ^{2}Ml}{2r^{2}},
\end{equation}%
and%
\begin{equation}
N_{k}^{2}(r)=k\log ((r^{2}-\kappa ^{2}Ml^{2}/2)/r_{0}^{2}),  \label{Nk}
\end{equation}%
where $k$ and $r_{0}$ are two real arbitrary constants. We use the
convention $\epsilon ^{tr\phi }=+1$. It is not hard to verify that (\ref%
{uncharged}) solves (\ref{caritaz}) in vacuum when $l\mu _{G}=1$.

That is, metric (\ref{uncharged}) represents an exact solution of
Topologically Massive Gravity that emerges at the chiral point. The Cotton
tensor associated to this solution is proportional to $k$, so that it is a
genuine solution to TMG in the sense that it does not solve Einstein
equation, except for the particular case $k=0$ where the metric becomes the
extremal BTZ black hole \cite{BTZ, BTZ2}. For all values of $k$ the metric
is clearly circularly symmetric and static, and thus compatible with $%
SO(2)\times \mathbb{R}$ symmetry.

In its ADM form, the metric reads%
\begin{equation}
ds^{2}=-\mathcal{N}_{\perp }^{2}(r)dt^{2}+\frac{dr^{2}}{N^{2}(r)}+\mathcal{R}%
^{2}(r)(d\phi -\mathcal{N}_{\phi }(r)dt)^{2},
\end{equation}%
where we have defined%
\begin{equation}
\mathcal{N}_{\perp }^{2}(r)=N^{2}(r)-r^{2}N_{\phi }^{2}(r)-N_{k}^{2}(r)+%
\mathcal{R}^{2}(r)\mathcal{N}_{\phi }^{2}(r),
\end{equation}%
and%
\begin{equation}
\mathcal{R}^{2}(r)=r^{2}+l^{2}N_{k}^{2}(r),\qquad \mathcal{N}_{\phi }(r)=%
\mathcal{R}^{-2}(r)(r^{2}N_{\phi }(r)+lN_{k}^{2}(r)).
\end{equation}

Metric (\ref{uncharged}) is actually nicely behaved. Despite the abstruse
form of the off-diagonal component $g_{\phi t}$, the determinant of the
metric is clearly $\det g=-r^{2}$, and the metric is Lorentzian for all
values of the radial coordinate $r$. The metric seems to present a horizon
at $r^{2}=$ $\kappa ^{2}Ml^{2}/2$. Nevertheless, for $k\neq 0$ the metric in
its form (\ref{uncharged}) is not defined for $r^{2}\leq $ $\kappa
^{2}Ml^{2}/2$ (for $k=0$ region $r^{2}<$ $\kappa ^{2}Ml^{2}/2$ would
correspond to the interior of the BTZ black hole). Let us analyze this
aspect together with the geodesic structure in more detail: At $r^{2}=$ $%
\kappa ^{2}Ml^{2}/2$, function $N_{k}^{2}$ diverges while $N^{2}$ vanishes.
Then, by analyzing the geodesic equation for massive particles, one observes
that the divergence of $N_{k}^{2}$ contributes to the radial effective
potential with a term like $\sim -(k/r^{2})\log (r^{2}-\kappa ^{2}Ml^{2}/2)$%
. This means that, for $k>0$, massive particles are scattered back when they
approach $r^{2}=$ $\kappa ^{2}Ml^{2}/2$, and this means that, at least for
positive $k$, the "horizon" is not actually there. In fact, for $k>0$ the
circle $r^{2}=\kappa ^{2}Ml^{2}/2$ turns out to be located at infinite
geodesic distance from any point. For $k<0$ the geodesic distance to a point
at $r^{2}=\kappa ^{2}Ml^{2}/2$ turns out to be finite. However, by taking a
look at the angular component of the geodesic equation one realizes that the
trajectories of massive particles wind indefinitely around the circle
defined by $r^{2}=$ $\kappa ^{2}Ml^{2}/2$ and thus these geodesic cannot be
extended across this circle \cite{Clementcomment}.

From (\ref{uncharged}) we also notice that $g_{tt}$ vanishes at $%
r^{2}=\kappa ^{2}Ml^{2}+kl^{2}\log ((r^{2}-\kappa ^{2}Ml^{2}/2)/r_{0}^{2})$,
and this always happens if $k\leq 0$. In particular, we know that for the
spinning BTZ (i.e. $k=0$) the radius $r=\kappa ^{2}Ml^{2}$ defines its
ergosphere \cite{BTZ2}. For $k>0$, however, metric function $g_{tt}$ only
vanishes if the parameters satisfy%
\begin{equation}
\kappa ^{2}M\geq 2k(1-\log (l^{2}k/r_{0}^{2})).  \label{ergos}
\end{equation}

For instance, let us consider the case $M=0$, for which the metric (\ref%
{uncharged}) takes the simple form%
\begin{equation}
ds^{2}=\frac{l^{2}}{r^{2}}dr^{2}+\frac{r^{2}}{l^{2}}(d\phi
^{2}-dt^{2})+k\log (\frac{r^{2}}{l^{2}})(dt-d\phi )^{2}=\frac{l^{2}}{r^{2}}%
dr^{2}+\frac{r^{2}}{l^{2}}dx^{+}dx^{-}+k\log (\frac{r^{2}}{l^{2}}%
)(dx^{-})^{2},  \label{inni}
\end{equation}%
where we defined $x^{\pm }=\phi \pm t$, we absorbed a factor $l$ in $\phi $,
and fixed $r_{0}$. From this expression we observe that if $k<0$ the
component $g_{tt}$ vanishes at $r^{2}=-2|k|l^{2}\log (r/r_{0})$, and that $%
g_{\phi \phi }$ may also vanish depending on $r_{0}$. On the other hand, if $%
k>0$ then the component $g_{\phi \phi }$ vanishes at $r^{2}=-2kl^{2}\log
(r/r_{0})$, and $g_{tt}$ may also vanish.

Now, let us move on and discuss the asymptotic behavior of (\ref{uncharged}%
). In the large $r$ limit, metric (\ref{uncharged}) takes the asymptotic form

\begin{eqnarray}
g_{tt} &=&-\frac{r^{2}}{l^{2}}+\mathcal{O}(\log (r))+\mathcal{O}(1),\qquad
g_{rr}=\frac{l^{2}}{r^{2}}+\mathcal{O}(r^{-4}),  \label{q1} \\
g_{\phi \phi } &=&r^{2}+\mathcal{O}(\log (r))+\mathcal{O}(1),\qquad g_{\phi
t}=\mathcal{O}(\log (r))+\mathcal{O}(1).  \label{q2}
\end{eqnarray}

We observe from this large $r$ expansion that this solution is not
asymptotically AdS$_{3}$ according to the definition given by Brown and
Henneaux in \cite{BH}. Nevertheless, (\ref{uncharged}) does still obey the
weakened AdS$_{3}$ asymptotic recently proposed by Grumiller and Johansson
in \cite{GJ,GJ2}. To see this, let us set $l=1$ for notational convenience,
and define the new coordinates $x^{\pm }=\phi \pm t$ and $y=r^{-1}$. In
terms of these coordinates, the large $r$ expansion of (\ref{uncharged})
reads

\begin{equation}
g_{--}=\mathcal{O}(\log (y))+\mathcal{O}(1),\qquad g_{-+}=y^{-2}+\mathcal{O}%
(1),\qquad g_{yy}=y^{-2}+\mathcal{O}(1),  \label{Nuevas}
\end{equation}%
together with $g_{++}=0$ and $g_{y\pm }=0$. It is worth noticing that
asymptotic behavior (\ref{Nuevas}) is strictly included in the boundary
conditions proposed in \cite{GJ,GJ2}, which, in addition, would also permit
a next-to-leading behavior like $g_{+y}=\mathcal{O}(y)$ and $g_{y-}=\mathcal{%
O}(y\log (y))$. These weakened boundary conditions were recently discussed
within the context of \textit{chiral gravity}, and these were shown to be
consistent with conformal asymptotic symmetry. In turn, this would permit to
define a consistent stress-tensor in the boundary. Our solution can be
thought of as a realization of the boundary conditions of \cite{GJ,GJ2}.

\subsection{Conserved charges and boundary terms}

Because the off-diagonal term in (\ref{uncharged}) grows logarithmically $%
\sim 2k\log (r)$ at large distance \cite{footnote}, it turns out that metric
(\ref{uncharged}) is not asymptotically AdS$_{3}$ in the sense of \cite{BH}.
However, we can still proceed to compute conserved charges of this solution
by holographic methods. After all, the solution is still asymptotically AdS$%
_{3}$ in the sense of the boundary conditions recently proposed in \cite%
{GJ,GJ2}. Then, we can resort to the method of defining an effective
stress-tensor induced on the boundary $\mathcal{\partial M}$, as in the case
of asymptotically locally AdS$_{3}$ solutions \cite{BK} (see also the
seminal paper \cite{BY}).

Consider the action with the boundary term,

\begin{equation}
I_{G}=\frac{1}{2\kappa ^{2}}\int_{\mathcal{M}}d^{3}x\sqrt{-g}(R+\frac{2}{%
l^{2}})+\frac{1}{\kappa ^{2}}\int_{\mathcal{\partial M}}d^{2}y\sqrt{-\gamma }%
K+\frac{1}{4\kappa ^{2}\mu _{G}}\int_{\mathcal{M}}d^{3}x\epsilon ^{\lambda
\mu \nu }\Gamma _{\lambda \sigma }^{\rho }(\partial _{\mu }\Gamma _{\nu \rho
}^{\sigma }+\frac{2}{3}\Gamma _{\mu \tau }^{\sigma }\Gamma _{\nu \rho
}^{\tau })  \label{ups}
\end{equation}%
where $K=$Tr$K=K_{i}^{i}$ is the trace of the extrinsic curvature $K_{ij}$.
Here, we see the Gibbons-Hawking term $\mathcal{B}$ appears. This action can
be expressed in terms of Gaussian coordinates $ds^{2}=d\eta ^{2}+\gamma
_{ij}dx^{i}dx^{j}$, with $K_{ij}=\frac{1}{2}\partial _{\eta }\gamma _{ij}$.
This reads \cite{KL,GJ} 
\begin{eqnarray}
I_{G} &=&\frac{1}{2\kappa ^{2}}\int_{\mathcal{M}}d^{2}y\text{ }d\eta \text{ }%
\sqrt{-\gamma }(R^{(2)}+K^{2}-\text{Tr}(K^{2})+\frac{2}{l^{2}})+  \notag \\
&+&\frac{1}{4\kappa ^{2}\mu _{G}}\int_{\mathcal{M}}d^{2}y\text{ }d\eta \text{
}\epsilon ^{ij}(-2K_{i}^{l}\partial _{\eta }K_{jl}+\Gamma _{in}^{l}\partial
_{\eta }\Gamma _{jl}^{n}+2K_{k}^{n}\Gamma _{in}^{l}\Gamma
_{jl}^{k}+K_{n}^{l}\partial _{j}\Gamma _{il}^{n}+\Gamma _{jn}^{l}\partial
_{i}K_{l}^{n})  \label{Sred}
\end{eqnarray}%
where Tr$K^{2}=K_{i}^{j}K_{j}^{i}$. In this expression, the Gibbons-Hawking
term does not appear because it cancels against a total derivative coming
from the bulk contribution. Expression (\ref{Sred}) turns out to be an
action for the metric $\gamma _{ij}$, which corresponds to the induced
metric in the boundary. The stress-tensor $T^{ij}$ associated to the
boundary manifold \cite{BY} is then obtained by varying (\ref{Sred}) with
respect to $\gamma _{ij}$ and evaluating it on-shell; namely, $\delta I_{G}=%
\frac{1}{2}\int_{\mathcal{\partial M}}d^{2}x\sqrt{-\gamma }T^{ij}\delta
\gamma _{ij}$. The conserved charges computed with this stress-tensor (see (%
\ref{carga}) below) diverge and then it is necessary to regularize the
action by adding an appropriate counter-term \cite{KL}. Such counter-term
turns out to be a cosmological constant term in the boundary; namely%
\begin{equation}
\Delta I_{G}=-\frac{1}{l\kappa ^{2}}\int d^{2}y\sqrt{-\gamma },
\label{LETAL}
\end{equation}%
which only depends on geometric quantities of the boundary, not affecting
the equations of motion in the bulk. A similar counter-term can be seen to
appear when analyzing other backgrounds of TMG. For instance, if one
considers the warped AdS$_{3}$ black hole of \cite{MCL, ALPSS}, the
counter-term in the boundary is also given by (\ref{LETAL}) but replacing
the overall factor $-1/l\kappa ^{2}$ by a $\mu _{G}$-dependent factor\ that
coincides with that in (\ref{LETAL}) when $\mu _{G}=3/l$ \cite{AGM}.

Including the counter-term (\ref{LETAL}), and in the case of asymptotically
AdS$_{3}$ spaces, the boundary stress-tensor takes the form

\begin{equation}
2\kappa ^{2}T^{ij}=2(K^{ij}-\gamma ^{ij}\text{Tr}K-\frac{1}{l}\gamma ^{ij})+%
\frac{1}{\mu _{G}}\epsilon ^{k(i}(\gamma ^{j)l}\partial _{\eta
}K_{kl}+2\partial _{\eta }K_{k}^{j)}).
\end{equation}

This expression can be used to compute conserved charges associated to
isometries on the boundary $\partial \mathcal{M}$. One is mainly concerned
with the conserved charges that are associated to Killing vectors $\partial
_{t}$\ and $\partial _{\phi }$, which correspond to the mass and the angular
momentum respectively. To define the charges it is convenient to make use of
the ADM formalism adapted to the boundary $\partial \mathcal{M}$. Then, the
charges are defined by \cite{BY}%
\begin{equation}
Q[\xi ]=\int ds\xi ^{i}u^{j}T_{ij},  \label{carga}
\end{equation}%
where $ds$ is the volume element of the constant-$t$ surfaces at the
boundary, $u$ is a unit vector orthogonal to the constant-$t$ surfaces, and $%
\xi $ is the Killing vector that generates the isometry in $\partial\mathcal{%
M}$.

To see how it works, let us consider the BTZ solution, whose metric is%
\begin{equation}
ds^{2}=-N^{2}(r)dt^{2}+\frac{dr^{2}}{N^{2}(r)}+r^{2}(d\phi +N_{\phi
}(r)dt)^{2}  \label{BTZ}
\end{equation}%
with 
\begin{equation}
N^{2}(r)=\frac{r^{2}}{l^{2}}-\kappa ^{2}M+\frac{\kappa ^{4}J^{2}}{4r^{2}}%
,\,\,\,\,\,\,\,\,\,\,\,\,\,\,\,\,N_{\phi }(r)=\frac{\kappa ^{2}J}{2r^{2}}%
,\,\,\,
\end{equation}

It is straightforward to compute the mass and the angular momentum of (\ref%
{BTZ}) following the recipe described above. The mass and the angular
momentum of BTZ black hole in TMG\ are then given by 
\begin{equation}
M_{\text{BTZ}}=M+\frac{J}{l^{2}\mu _{G}},\qquad \qquad J_{\text{BTZ}}=J+%
\frac{M}{\mu _{G}},
\end{equation}%
respectively. It is well known \cite{MCL,KL} that this result differs from
the charges of the same solution for GR, which are recovered if $1/\mu
_{G}=0 $. In particular, these values for the mass and angular momentum in
TMG imply that at the chiral point $\mu _{G}=1/l$ all the BTZ black holes in
TMG\ fulfill the relation $J_{\text{BTZ}}=lM_{\text{BTZ}}=lM+J$. More
specifically, if $J=-lM$ at the chiral point both the mass and the angular
momentum vanish.

Then, we can use the same idea to compute the mass and angular momentum of (%
\ref{uncharged}). It yields%
\begin{equation}
M_{(k)}=\frac{6\pi k}{\kappa ^{2}},\qquad \qquad J_{(k)}=-\frac{6\pi lk}{%
\kappa ^{2}},  \label{kargas}
\end{equation}

This is consistent with the fact that (\ref{uncharged}) is a perturbation of
the extremal BTZ\ black hole with $J=-lM$ at the chiral point $\mu _{G}=1/l$%
. Recall that BTZ\ black hole with bare parameters obeying $J=-lM$ in chiral
gravity have zero mass and zero angular momentum, and then we interpret it
as the \textit{ground state} for (\ref{uncharged}). Notice that, as long as
Newton constant is positive, the BTZ black hole in TMG\ have positive mass,
and our solution (\ref{uncharged}) has also positive mass for $k>0$.
Conversely, if we adopt the \textit{wrong sign} for Newton constant (what
amounts to change $\kappa ^{2}\rightarrow -\kappa ^{2}$ in (\ref{Action})
but keeping $\kappa ^{2}M$ unchanged) then the BTZ black hole turns out to
have negative mass, while (\ref{uncharged}) has positive mass for $k<0$.

Before concluding this section, let us mention that at the point $l\mu
_{G}=-1$ one also finds a vacuum solution of TMG with the form%
\begin{equation}
ds^{2}=-N^{2}(r)dt^{2}+\frac{dr^{2}}{N^{2}(r)}+r^{2}(N_{\phi }(r)dt+d\phi
)^{2}+N_{k}^{2}(r)(r^{2}-\kappa ^{2}Ml^{2}/2)(dt+ld\phi )^{2}.
\label{uncharged2}
\end{equation}

Unlike solution (\ref{uncharged}), this metric tends to that of the extremal
BTZ\ black hole when $r$ approaches the horizon $r^{2}=\kappa ^{2}Ml^{2}/2$.
The off-diagonal term in (\ref{uncharged2}), however, grows in more drastic
way, behaving like $\sim 2kr^{2}\log r$ at large distances.

Also, a charged solution at the chiral point exists, and it has a form like (%
\ref{uncharged}) and (\ref{uncharged2}) with its charge associated to $k$.
Now, we move on to discuss charged solutions.

\section{Charged solutions with a Chern-Simons term}

\subsection{The solutions}

In this section, we will show that solution (\ref{uncharged}) admits a
natural generalization when TMG is coupled to TME (\ref{A2}) if the coupling
constants satisfy 
\begin{equation}
l\mu _{G}=1+2l\mu _{E}.  \label{punto}
\end{equation}

For further convenience, we define the parameter $\varepsilon =-l\mu _{E}=%
\frac{1}{2}(1-l\mu _{G})$, which is an arbitrary real number. In particular,
the theory at the chiral point corresponds to $\varepsilon =0$ and $%
\varepsilon =1$. For the case $l\mu _{G}=1+2l\mu _{E}=1-2\varepsilon >1$,
the metric of the charged solution takes the form

\begin{equation}
ds^{2}=-N^{2}(r)dt^{2}+\frac{dr^{2}}{N^{2}(r)}+r^{2}(d\phi -N_{\phi
}(r)dt)^{2}-N_{Q}^{2}(r)(dt-ld\phi )^{2}  \label{charged}
\end{equation}%
with%
\begin{equation*}
N^{2}(r)=\frac{r^{2}}{l^{2}}-\kappa ^{2}M+\frac{\kappa ^{4}M^{2}l^{2}}{4r^{2}%
},\qquad \qquad N_{\phi }(r)=\frac{\kappa ^{2}Ml}{2r^{2}},
\end{equation*}%
and with 
\begin{equation*}
N_{Q}^{2}(r)=\frac{1}{2}\kappa ^{2}Q^{2}(r^{2}-\kappa ^{2}Ml^{2}/2)^{-l\mu
_{E}}\log (\left( r^{2}-\kappa ^{2}Ml^{2}/2\right) /r_{0}^{2}),
\end{equation*}%
and the electromagnetic field takes the form%
\begin{equation}
A(r)=A_{0}\left( r^{2}-\kappa ^{2}Ml^{2}/2\right) )^{-l\mu _{E}/2}(dt+ld\phi
),\qquad A_{0}^{2}=Q^{2}\frac{2(l\mu _{E}+1)}{l\mu _{E}(2l\mu _{E}+1)}.
\label{A}
\end{equation}

Again, metric (\ref{charged}) corresponds to a deformation of the extremal
BTZ black hole, which corresponds to the uncharged case $Q=0$. If $l\mu
_{E}>0$, function $N_{Q}^{2}$ in (\ref{charged}) diverges at $r^{2}=\kappa
^{2}Ml^{2}/2$, but the curvature invariants remain constant. In fact, for
all the solutions we find the Ricci scalar%
\begin{equation}
R=-\frac{6}{l^{2}},  \label{Ricci}
\end{equation}%
and the Kretschmann scalar,%
\begin{equation}
R_{\mu \nu \rho \sigma }R^{\mu \nu \rho \sigma }=R_{\mu \nu }R^{\mu \nu }=%
\frac{12}{l^{4}};  \label{Kretschmann}
\end{equation}%
and one also finds $C_{\mu \nu }C^{\mu \nu }=0$.

It is worth noticing that (\ref{Ricci}) holds even for charged solutions.
This implies that the gauge field configuration is such that $T_{\mu }^{\mu
}=0$. We will return to this point below. It is also interesting that the
Kretschmann scalar turns out to be independent of the parameters of the
solution $Q$ and $M$. This is a curious fact since solutions of Einstein
gravity coupled to matter yielding traceless stress-tensor generically
depend on the integration constants of the solution \cite{MZ}. The fact that
both the Ricci and Kretschmann scalars take the same value for all the
members of the family of metrics (\ref{charged}) could lead to suspect that
all these geometries correspond to discrete quotients of the same (vacuum)
space. However, this cannot be the case for $all$ the solutions since the
case $Q=0$ (resp $k=0$ in (\ref{uncharged})) is locally AdS$_{3}$ while $%
Q\neq 0$ has non vanishing Cotton tensor.

The asymptotic behavior of (\ref{charged}) is determined by the following
expansion

\begin{eqnarray*}
g_{tt} &=&-\frac{r^{2}}{l^{2}}+\mathcal{O}(1)+\mathcal{O}(r^{-2l\mu
_{E}}\log (r)),\qquad g_{rr}=\frac{l^{2}}{r^{2}}+\mathcal{O}(r^{-4}), \\
g_{\phi \phi } &=&r^{2}+\mathcal{O}(1)+\mathcal{O}(r^{-2l\mu _{E}}\log
(r)),\qquad g_{\phi t}=\mathcal{O}(1)+\mathcal{O}(r^{-2l\mu _{E}}\log (r)),
\end{eqnarray*}%
and $g_{r\phi }=g_{rt}=0$. That is, solutions (\ref{uncharged}) are
asymptotically AdS$_{3}$ for $l\mu _{E}=-\varepsilon >0$. However, for $l\mu
_{E}>0$, the gauge field (\ref{A}) diverges at $r^{2}=\kappa ^{2}Ml^{2}/2$.
It turns out that solutions for which the gauge field vanishes at the
"horizon" (e.g. for $l\mu _{E}<0$) diverges dramatically at the boundary,
and viceversa, and thus no hair is allowed in this sense.

\subsection{The logarithmic branch of self-dual solutions}

As mentioned, the fact that the Ricci scalar of (\ref{charged}) takes the
value $R=-6l^{-2}$ tells us that these charged solutions satisfy the
traceless condition $T_{\mu }^{\mu }=0$, which in three-dimensions implies $%
F_{\mu \nu }F^{\mu \nu }=0$. This is reminiscent of the self-dual solutions
discussed by Ait Moussa and Cl\'{e}ment in \cite{CM}. Then, a natural
question is whether our solutions are somehow related to those of \cite{CM}.
We will see that, even though solutions (\ref{charged}) were not considered
in \cite{CM}, these can be obtained starting from the ones considered in
that paper by taking the limit $l\mu _{G}\rightarrow 1+2l\mu _{E}$
appropriately, and then extending the manifold. To see this, let us define
the coordinate 
\begin{equation*}
\frac{2}{\mu _{E}}\rho =r^{2}-\frac{\kappa ^{2}Ml^{2}}{2}.
\end{equation*}

In terms of the new radial coordinate $\rho $ the metric (\ref{charged})
takes the form%
\begin{eqnarray}
ds^{2} &=&-(\frac{2}{l^{2}\mu _{E}}\rho +\frac{1}{2}\mathcal{Q}^{2}\rho
^{-l\mu _{E}}\log \left( \rho \right) -\frac{1}{2}\mathcal{M})dt^{2}-l(%
\mathcal{Q}^{2}\rho ^{-l\mu _{E}}\log (\rho )-\mathcal{M})dtd\phi +  \notag
\\
&&+\frac{l^{2}d\rho ^{2}}{4\rho ^{2}}+l^{2}(\frac{2}{l^{2}\mu _{E}}\rho -%
\frac{1}{2}\mathcal{Q}^{2}\rho ^{-l\mu _{E}}\log \left( \rho \right) +\frac{1%
}{2}\mathcal{M})d\phi ^{2},  \label{CM0}
\end{eqnarray}%
where we defined%
\begin{equation*}
\mathcal{Q}^{2}=(\mu _{E}/2)^{l\mu _{E}}\kappa ^{2}Q^{2},\qquad \mathcal{M}%
=\kappa ^{2}M.
\end{equation*}

In \cite{CM}, similar solutions were considered for the case $l\mu
_{G}-1\neq 2l\mu _{E}$, and these have the slightly different form%
\begin{eqnarray}
ds^{2} &=&-(\frac{2}{l^{2}\mu _{E}}\rho +\frac{1}{2}\mathcal{Q}^{2}\rho
^{-l\mu _{E}}+\frac{1}{2}\mathcal{J}\rho ^{(1-l\mu _{G})/2}-\frac{1}{2}%
\mathcal{M})dt^{2}-l(\mathcal{Q}^{2}\rho ^{-l\mu _{E}}+\mathcal{J}\rho
^{(1-l\mu _{G})/2}-\mathcal{M})dtd\phi +  \notag \\
&&+\frac{l^{2}d\rho ^{2}}{4\rho ^{2}}+l^{2}(\frac{2}{l^{2}\mu _{E}}\rho -%
\frac{1}{2}\mathcal{Q}^{2}\rho ^{-l\mu _{E}}-\frac{1}{2}\mathcal{J}\rho
^{(1-l\mu _{G})/2}+\frac{1}{2}\mathcal{M})d\phi ^{2}.  \label{CM}
\end{eqnarray}

Therefore, solutions (\ref{charged}) arise in the limit $l\mu
_{G}\rightarrow 1+2l\mu _{E}$ of (\ref{CM}). At the point (\ref{punto}), two
independent solutions to the field equations degenerate and thus the
logarithmic form $\sim \mathcal{Q}^{2}\rho ^{-l\mu _{E}}\log \rho $ stands
as a new linear independent solution. The other solution $\sim \mathcal{J}%
\rho ^{-l\mu _{E}}=\mathcal{J}\rho ^{(1-l\mu _{G})/2}$ contributes by
setting the scale $\rho _{0}$ (related to $r_{0}$ in (\ref{charged})) where
the logarithm vanishes.

The case (\ref{CM}) we consider here is somehow special. It is continuously
connected to the vacuum solutions (\ref{uncharged}) and (\ref{uncharged2}).
Likely, solution (\ref{uncharged}) can be associated to a particular limit
of solutions studied in \cite{C2}.

Notice that for $Q=0$ the region $-l^{2}\mu _{E}\kappa ^{2}M/4\leqslant \rho
<0$ corresponds to the region inside the horizon of the extremal BTZ black
hole. Recall that for $k<0$ the point $\rho =0$ is at finite geodesic
distance from any point located at $\rho >0$, and the geodesics end there.
On the other hand, for $k>0$ the point $\rho =0$ is at infinite geodesic
distance, as it happens for the self-dual solutions considered in \cite{CM}.

\subsection{Reduced field equations}

The relation with the self-dual solutions (\ref{CM}) suggests that we could
use the techniques used \cite{CM} to rederive our solutions (\ref{CM0}). The
idea in \cite{CM} was to reduce the field equations of TMG to a relativistic
dynamical system which is easily solved by choosing the appropriate ansatz.

Consider with \cite{CM} the following parameterization of the metric

\begin{equation}
ds^{2}=h_{ab}(\rho )dx^{a}dx^{b}+\frac{1}{\zeta ^{2}R^{2}(\rho )}d\rho
^{2},\qquad A(\rho )=\psi _{a}(\rho )dx^{a},
\end{equation}%
where $a,b=0,1$, with $x^{0}=t$, \ $x^{1}=\varphi $, and $h_{tt}+h_{\phi
\phi }=2T$, $h_{tt}-h_{\phi \phi }=2X$, $h_{\phi t}=Y$. Here, $X^{0}=T$, $%
X^{1}=X$, and $X^{2}=Y$, are functions of $\rho $ that satisfy the Minkowski
product $R^{2}=\mathbf{X}^{2}=-T^{2}+X^{2}+Y^{2}=\eta _{ij}X^{i}X^{j}$ with $%
i,j=0,1,2$. We also denoted $\psi _{0}=A_{t}$ and $\psi _{1}=A_{\phi }$ for
convenience.

In terms of this variables, the action takes the form%
\begin{equation}
I_{G}+I_{E}=\frac{1}{2}\int d^{2}x\int d\rho \left( \frac{1}{2\kappa ^{2}}%
\zeta \dot{X}_{i}\dot{X}^{i}+\frac{2}{\kappa ^{2}l^{2}}\zeta ^{-1}+\frac{1}{%
2\kappa ^{2}\mu _{G}}\zeta ^{2}\varepsilon _{ijk}X^{i}\dot{X}^{j}\ddot{X}%
^{k}+\zeta \dot{\psi}\Sigma ^{0}\Sigma _{i}X^{i}\dot{\psi}-\mu _{E}\psi
\Sigma ^{0}\dot{\psi}\right)  \label{300Themovie}
\end{equation}%
where $\mathbf{X}=(X^{0},X^{1},X^{2})$, $X_{i}=\eta _{ij}X^{j}$ with $%
i,j,k=0,1,2$. We also denoted $\mathbf{\dot{X}}=\partial _{\rho }\mathbf{X}$%
, $\mathbf{\ddot{X}}=\partial _{\rho }^{2}\mathbf{X}$, etc. The components
of the vector $\mathbf{\Sigma }=(\Sigma ^{0},\Sigma ^{1},\Sigma ^{2})$ are
given by $\Sigma ^{0}=\sigma ^{1}$, $\Sigma ^{0}=i\sigma ^{2}\sigma ^{1}$,
and $\Sigma ^{2}=\sigma ^{3}$, where $\sigma ^{i}$ are the Pauli matrices
acting on the two-component vectors $\psi =(\psi _{0},\psi _{1})$. Notice
also that $\zeta $ stands in (\ref{300Themovie}) as a Lagrange multiplier.
The first two terms in the action above correspond to the Einstein-Hilbert
term and the cosmological constant term, while the third one corresponds to
the gravitational Chern-Simons term. On the other hand, the terms involving $%
\psi $ come from the gauge field $A=\psi _{a}dx^{a}$ with $a=0,1$.

The products in (\ref{300Themovie}) are defined as $\mathbf{X}\cdot \mathbf{Y%
}=X^{i}Y^{j}\eta _{ij}$, $(\mathbf{X}\wedge \mathbf{Y})^{k}=\eta
^{kl}\varepsilon _{ijl}X^{i}Y^{j}$, and then the action can be written as
follows

\begin{equation}
I_{G}+I_{E}=\frac{1}{2}\int d^{2}x\int d\rho \left( \frac{1}{2\kappa ^{2}\mu
_{G}}\zeta ^{2}\mathbf{X}\cdot (\mathbf{\dot{X}}\wedge \mathbf{\ddot{X}})+%
\frac{1}{2\kappa ^{2}}\zeta \mathbf{\dot{X}}^{2}+\zeta \dot{\psi}\Sigma ^{0}%
\mathbf{\Sigma }\cdot \mathbf{X}\dot{\psi}-\mu _{E}\psi \Sigma ^{0}\dot{\psi}%
+\frac{2}{\kappa ^{2}l^{2}}\zeta ^{-1}\right) ,  \label{30}
\end{equation}

Varying this action with respect to $\psi $, we find

\begin{equation}
\frac{\partial }{\partial \rho }(\zeta (\mathbf{\Sigma }\cdot \mathbf{X})%
\dot{\psi}+\mu _{E}\psi )=0.  \label{3}
\end{equation}%
This equation yields%
\begin{equation}
\zeta \mathbf{\dot{S}}_{E}=\frac{2\mu _{E}}{R^{2}}\mathbf{X}\wedge \mathbf{S}%
_{E},\qquad \text{with \qquad }\mathbf{S}_{E}=-\frac{\kappa ^{2}}{2}\bar{\psi%
}\mathbf{\Sigma }\psi  \label{4}
\end{equation}

Also, varying (\ref{30}) with respect to $\mathbf{X}$ we find%
\begin{equation}
\mathbf{\ddot{X}}=\frac{\zeta }{2\mu _{G}}(3(\mathbf{\dot{X}}\wedge \mathbf{%
\ddot{X}})+2(\mathbf{X}\wedge \mathbf{\dddot{X}}))-\frac{2\mu _{E}^{2}}{%
\zeta ^{2}R^{2}}(\mathbf{S}_{E}-\frac{2}{R^{2}}\mathbf{X}(\mathbf{S}%
_{E}\cdot \mathbf{X}))  \label{5}
\end{equation}%
and 
\begin{equation}
\mathbf{S}_{E}\cdot \mathbf{X}=\frac{\zeta ^{2}R^{2}}{2\mu _{E}^{2}}(\mathbf{%
X}\cdot \mathbf{\ddot{X}}-\frac{3\zeta }{2\mu _{G}}\mathbf{X}\cdot (\mathbf{%
\dot{X}}\wedge \mathbf{\ddot{X}})).  \label{6}
\end{equation}

The Hamiltonian constraint comes from varying (\ref{30}) with respect to $%
\zeta $,

\begin{equation}
\mathcal{H}=\frac{1}{4\kappa ^{2}}(\mathbf{\dot{X}}^{2}+2\mathbf{X}\cdot 
\mathbf{\ddot{X}}-\frac{\zeta }{\mu _{G}}\mathbf{X}\cdot (\mathbf{\dot{X}}%
\wedge \mathbf{\ddot{X}})-\frac{4}{l^{2}\zeta ^{2}})=0.  \label{10}
\end{equation}

Now, let us look for solutions of the form%
\begin{equation}
\mathbf{X}(\rho )=\mathbf{u}G(\rho )+\mathbf{v}F(\rho )  \label{1}
\end{equation}%
where $F$ and $G$ are functions of $\rho $, while $\mathbf{u}$ and $\mathbf{v%
}$ are two vectors such that $\mathbf{u}\cdot \mathbf{v}=\eta
_{ij}u^{i}v^{j}=0$, and $\mathbf{v}^{2}=\eta _{ij}v^{i}v^{j}=0$. This
implies $\mathbf{u}\wedge \mathbf{v}=\lambda \mathbf{v}$, that is $\eta
^{kl}\varepsilon _{ijl}u^{i}v^{j}=\lambda v^{k}$, where $\lambda $ is an
arbitrary constant. We can make the choice \cite{unidades}

\begin{equation}
\mathbf{u}=\frac{1}{2l}(1-l^{2},1+l^{2},0),\qquad \mathbf{v}=-\frac{1}{4}%
(1+l^{2},1-l^{2},\mp 2l)  \label{uv}
\end{equation}%
and then $\mathbf{u}^{2}=\eta _{ij}u^{i}u^{j}=1$. Then, we have two possible
choices for $\lambda $, namely $\lambda =\pm 1,$ which correspond to each
possibility for the sign $\pm $ in (\ref{uv}). This ambiguity in the sign
will be ultimately related to the sign of $l\mu _{G}$.

In terms of the ansatz (\ref{1})-(\ref{uv}), the Hamiltonian constraint (\ref%
{10}) reads

\begin{equation}
\overset{.}{G}^{2}+2G\overset{...}{G}-\frac{4}{^{l^{2}\mu _{E}^{2}}}=0.
\label{10n}
\end{equation}

On the other hand, the equations of motion give

\begin{equation}
\mathbf{S}_{E}=\frac{\lambda \mu _{E}}{4\mu _{G}}G^{2}(-3\overset{.}{F}%
\overset{..}{G}+3\overset{.}{G}\overset{..}{F}+2G\overset{...}{F}-2F\overset{%
...}{G})\mathbf{v}+\left( G\mathbf{u}+F\mathbf{v}\right) G\overset{..}{G}-%
\frac{1}{2}(\overset{..}{G}\mathbf{u}+\overset{..}{F}\mathbf{v})G^{2}.
\label{EstA}
\end{equation}

Varying with respect to $\rho $ we obtain $\mathbf{\dot{S}}_{E}$. We find $%
\mathbf{X}\wedge \mathbf{S}_{E}$ with $R^{2}=G^{2}$, and we can go back to (%
\ref{4}) and find%
\begin{equation*}
\mathbf{\dot{S}}_{E}=\frac{2}{R^{2}}\mathbf{X}\wedge \mathbf{S}_{E},
\end{equation*}%
with $\zeta =\mu _{E}$.

Then, we can solve both for $\mathbf{u}$ and for $\mathbf{v}$. Before doing
this, let us further specify the ansatz%
\begin{equation}
G(\rho )=a\rho \text{ \ \ \ \ \ \ \ \ \ }F(\rho )=\kappa ^{2}Q^{2}\rho
^{\varepsilon }\ln \rho -\kappa ^{2}M  \label{13}
\end{equation}%
where $Q$ and $M$ are two arbitrary real constants, while $a$ and $%
\varepsilon $ are two real parameters to be determined. This ansatz
automatically satisfies the equation for $\mathbf{u}$, and thus we only have
to solve for $\mathbf{v}$. The Hamiltonian constraint (\ref{10}) demands

\begin{equation}
a^{2}=-\frac{4\Lambda }{\mu _{E}^{2}}=\frac{4}{l^{2}\mu _{E}^{2}}  \label{14}
\end{equation}%
which is only possible if $\Lambda =-l^{-2}<0$. Then, we have $a=\pm 2/l\mu
_{E}$.

On the other hand, from the equation for $\mathbf{v}$ we find

\begin{equation}
\frac{\mu _{E}}{\mu _{G}}a^{2}\left( \varepsilon -\frac{1}{2}\right) \left( 
\frac{1}{2}\varepsilon \lambda a-1\right) \varepsilon (\varepsilon
-1)=a\left( \frac{\varepsilon }{2}a-\lambda \right) \varepsilon (\varepsilon
-1),  \label{16}
\end{equation}

and%
\begin{equation}
\frac{\mu _{E}}{\mu _{G}}a^{3}\lambda \varepsilon \left( 2\varepsilon ^{2}-%
\frac{9}{4}\varepsilon +\frac{1}{2}\right) +a^{2}\varepsilon \left( -\frac{3%
}{2}\varepsilon +1\right) =\frac{\mu _{E}}{\mu _{G}}a^{2}\left( 3\varepsilon
^{2}-3\varepsilon +\frac{1}{2}\right) -a\left( 2\varepsilon -1\right)
\lambda .  \label{17}
\end{equation}

Let us first analyze the cases $0\neq \varepsilon \neq 1$. From (\ref{16})
we find that $\lambda =\pm 1$ and $a=\pm 2/\varepsilon $, what implies $l\mu
_{E}=\pm \varepsilon $. Then, from both (\ref{16}) and\ (\ref{17}) we get $%
\mu _{G}/\mu _{E}=(2\varepsilon -1)/\varepsilon $. That is,%
\begin{equation}
l\mu _{G}=2l\mu _{E}\mp 1.  \label{23}
\end{equation}

For these cases $\varepsilon =0$ and $\varepsilon =1$, equation (\ref{16})
is trivially satisfied and we get no restriction for $a$ and $\mu _{E}$. For 
$\varepsilon =0$ equation (\ref{17}) yields $l\mu _{G}=1$, which corresponds
to the chiral point. In fact, $\varepsilon =0$ corresponds to the solution (%
\ref{uncharged}) since for this configuration we also find $\psi _{0}=\psi
_{1}=0$, so that%
\begin{equation}
A_{\mu }=0.
\end{equation}

Similarly, for $\varepsilon =1$ equation (\ref{17}) implies $l\mu _{G}=-1$
with $\psi _{t}=\psi _{\varphi }=0$, and this is solution (\ref{uncharged2}).

For the generic case, the metric takes the form%
\begin{eqnarray}
ds^{2} &=&\left( \pm \frac{2}{l^{2}\mu _{E}}\rho -\frac{1}{2}\kappa
^{2}Q^{2}\rho ^{\pm l\mu _{E}}\log \left( \rho \right) +\frac{1}{2}\kappa
^{2}M\right) dt^{2}+l\left( \kappa ^{2}Q^{2}\rho ^{\eta }\log \left( \rho
\right) -\kappa ^{2}M\right) dtd\varphi  \notag \\
&&-l^{2}\left( \pm \frac{2}{l^{2}\mu _{E}}\rho +\frac{1}{2}\kappa
^{2}Q^{2}\rho ^{\pm l\mu _{E}}\log \left( \rho \right) -\frac{1}{2}\kappa
^{2}M\right) d\varphi ^{2}+\frac{l^{2}}{4\rho ^{2}}d\rho ^{2}  \label{27}
\end{eqnarray}%
and the gauge field configuration is%
\begin{equation}
A(\rho )=\sqrt{\frac{2Q^{2}(1-\varepsilon )\rho ^{\varepsilon }}{\varepsilon
(2\varepsilon -1)}}(dt+ld\phi )=Q\sqrt{\frac{2(1\mp l\mu _{E})}{l\mu
_{E}(2l\mu _{E}\mp 1)}}\rho ^{\pm l\mu _{E}/2}(dt+ld\phi ).  \label{33}
\end{equation}

Expressions (\ref{27})-(\ref{33}) correspond to solutions (\ref{charged}).
Thus, we have rederived solutions (\ref{charged}) by using the method of 
\cite{CM}. This method also permits to compute conserved charges of the
solutions in a rather systematic way. This amounts to calculate the so
called super-angular momentum $\mathbf{J}$, which is a current that gathers
the conserved charges of this type of background with two commuting Killing
vectors. The expression for such current is%
\begin{equation*}
\mathbf{J}=\mathbf{L}+\mathbf{S}_{G}+\mathbf{S}_{E}
\end{equation*}%
where%
\begin{equation*}
\mathbf{L}=\frac{1}{2\kappa ^{2}}\mathbf{X}\wedge \mathbf{\dot{X}},\qquad 
\mathbf{S}_{G}=\frac{1}{4\kappa ^{2}\mu _{G}}\left( 2\mathbf{X}\wedge (%
\mathbf{X}\wedge \mathbf{\ddot{X}})-\mathbf{\dot{X}}\wedge (\mathbf{X}\wedge 
\mathbf{\dot{X}})\right) .
\end{equation*}%
Evaluated in (\ref{27}), these take the form%
\begin{eqnarray*}
\mathbf{L} &=&\frac{\lambda a}{2}(Q^{2}(\varepsilon -1)\rho ^{\varepsilon
}\log \rho +Q^{2}\rho ^{\varepsilon }+M)\mathbf{v}, \\
\mathbf{S}_{G} &=&\frac{a^{2}}{4\mu _{G}}\left( (2\varepsilon
-1)(\varepsilon -1)Q^{2}\rho ^{\varepsilon }\log \rho +(4\varepsilon
-3)Q^{2}\rho ^{\varepsilon }-M\right) \mathbf{v},
\end{eqnarray*}%
and, from (\ref{EstA}), we can also find the expression for $\mathbf{S}_{E}$%
. For the vacuum solution (\ref{uncharged}), which corresponds to the case $%
\varepsilon =0$, we find%
\begin{equation}
\mathbf{J}=-\frac{2k}{l\kappa ^{2}}\mathbf{v},  \label{Jota}
\end{equation}%
where we have set $a$ to take a convenient value. Notice that (\ref{Jota})
turns out to be proportional to $k/\kappa ^{2}$, like in (\ref{kargas}). In 
\cite{BC} the computation of conserved charges from the expression for $%
\mathbf{J}$ is discussed in detail, and the mass and angular momentum can be
computed as quantities associated to Killing vectors $\partial _{t}$ and $%
\partial _{\phi }$ respectively. Remarkably, the mass and angular momentum
computed with this method agree with our result (\ref{kargas}), which was
calculated by considering the stress-tensor in the boundary \cite%
{Clementcomment2}. We will not give the details of this computation here;
instead, we draw the reader's attention to the very interesting papers \cite%
{CM,C2,MCL} and \cite{Clementnuevo}.

\section{Summary}

We have studied solutions to Cosmological Topologically Massive Gravity at
special values of the coupling constants. First, we considered the theory at
the chiral point, for which vacuum solution (\ref{uncharged}) was exhibited.
This solution corresponds to a one-parameter deformation of GR solutions and
is continuously connected to the extremal BTZ black hole. To be more
precise, solution (\ref{uncharged}) has two parameters, $k$ and $M$, and
when $k=0$ the solution turns out to be the extremal BTZ black hole with
bare parameters satisfying $J=-lM$. It is well known that for all values of $%
J$ and $M$, the BTZ black holes in TMG at the chiral point \cite{LSS}
satisfy the extremality relation $J_{\text{BTZ}}=lM_{\text{BTZ}}=lM+J$. In
turn, the (massless) extremal BTZ with $lM+J=0$ can be thought of as a kind
of \textit{ground state} of solutions (\ref{uncharged}), which are labeled
by a real number $k$.

Solution (\ref{uncharged}) fails to be asymptotically AdS$_{3}$ in the sense
of Brown-Henneaux boundary conditions \cite{BH}, and this is because of a
logarithmic damping at large distances. Nevertheless, it is still
asymptotically AdS$_{3}$ in the sense of the boundary conditions recently
proposed by Grumiller and Johansson in \cite{GJ,GJ2}. Then, the holographic
computation of conserved charges in terms of the boundary stress-tensor
yielded (\ref{kargas}), and the mass and angular momentum turn out to be
proportional to $k/\kappa ^{2}$. Therefore, the sign of the mass of (\ref%
{uncharged}) can be chosen to be opposite to that of the BTZ black hole in
this theory. That is, if one adopts the negative sign for the Newton
constant (as it is usual in TMG) then the solutions with positive mass ($k<0$%
) are those that allow geodesic to reach the radius $r^{2}=\kappa
^{2}Ml^{2}/2$ at finite proper time, while for the case $k>0$ that circle is
at infinite geodesic distance.

We also considered solutions (\ref{charged}), which are charged analogues to
(\ref{uncharged}) that exist when the coupling constants satisfy the
relation $l\mu _{G}=1+2l\mu _{E}$. Unlike vacuum solutions we found at the
chiral point, their charged analogues may have a stronger damping at large
distance and then represent asymptotically AdS$_{3}$ solutions in the sense
of \cite{BH}. However, for asymptotically AdS$_{3}$ solutions both the gauge
field and the effective potential of the geodesic equation for massive
particles diverge at the horizon.

Like vacuum solutions, charged solutions (\ref{charged}) have constant
scalar curvature $R=-6l^{-2}$. This implies that the corresponding gauge
field configurations fulfill the condition $T_{\mu }^{\mu }=F_{\mu \nu
}F^{\mu \nu }=0$. \ This is reminiscent of the self-dual solutions studied
in reference \cite{CM}, and we have actually shown that solutions (\ref%
{charged}) can be thought of as a limiting procedure starting from the
self-dual solutions of \cite{CM}.

It is also remarkable that both solutions (\ref{uncharged}) and (\ref%
{charged}) have constant Kretschmann scalar, given by $R_{\mu \nu \sigma
\rho }R^{\mu \nu \sigma \rho }=12l^{-4}$. This means that all the quadratic
invariants turn out to be independent of the two parameters of the
solutions. Nevertheless, it is worth emphasizing that both $M$ and $Q$ still
represent actual parameters labeling the solutions, as they enter in the
computation of the charges in a non trivial way, and, besides, parameter $Q$
is the one that permits to interpolate between (\ref{charged}) and the
extremal BTZ black hole.

Before concluding, let us comment on the relation between the solutions we
discussed here and a class of $pp$-wave solutions recently discussed in the
literature. Just recently, we were taught \cite{Troncosocomment} that
solution (\ref{uncharged}) can be obtained from one of the $pp$-wave
solutions considered in \cite{Gibbons} by an appropriate coordinate
transformation, in addition to the compactification of the direction we
denoted by $\phi $. The metrics considered in \cite{Gibbons} have the form%
\begin{equation}
ds^{2}=dR^{2}+e^{2R}dx^{+}dx^{-}+R\text{ }f(x^{-})(dx^{-})^{2},
\label{gibbons}
\end{equation}%
where $f(x^{-})$ is an arbitrary function of $x^{-}$ (see Eq. (3.21) of Ref. 
\cite{Gibbons}, with $l=1$, $\mu =1$, $\rho =R$, $u=x^{+}/2$, $v=x^{-}$, and 
$x^{-}=t-\phi $, with $\phi $ being compact). It is easy to see that (\ref%
{gibbons}) can be written as our solution (\ref{uncharged}) by means of the
appropriate coordinate transformation. For instance, consider the case $M=0$%
, as in (\ref{inni}), which takes the form (\ref{gibbons}) by choosing $f=2k$
and defining the radial coordinate $R=\log (r)$. A similar relation holds
between (\ref{uncharged2}) and the solution presented in Eq. (3.22) of \cite%
{Gibbons} for the case $l\mu _{G}=-1$. This allows to interpret our charged
solutions (\ref{charged}) as a generalization of some of the solutions
considered in \cite{Gibbons}.

The solution we have presented here generalizes the extremal BTZ black hole
solution at the chiral point, and it represents an exact realization of the
boundary conditions proposed in \cite{GJ,GJ2}.

\section{Addendum: A persistent solution of the New Massive Gravity}

A few months ago, after this paper was published, a new theory of massive
gravity in three dimensions was presented \cite{NMG}. This theory is defined
by supplementing the Einstein-Hilbert action with particular
quadratic-curvature corrections, which, at linearized level, turn out to
coincide with the Fierz-Pauli Lagrangian for a massive spin-2 particle. This
new theory received the name of New Massive Gravity (NMG), to be
distinguished from the Topologically Massive Gravity. It was shown in \cite%
{Sun} that adding the Lagrangian of the NMG to that of the TMG also leads to
very special AdS$_{3}$ asymptotic at the particular point of the moduli
space where one of the central charges of the dual conformal field theory
vanishes. This and other analogies suggest that one could also have the
Chiral Gravity vs. Log Gravity dichotomy \cite{Maloney} in the model that
includes both NMG and TMG Lagrangians. What we show in this \textit{addendum}
is that the uncharged solution we found in Section 2 for the case of Log
Gravity actually persists as an exact solution in the generalized massive
gravity model defined by including the higher-curvature terms.

Let us consider the action 
\begin{equation*}
S=\frac{1}{2\kappa ^{2}}\int d^{3}x\sqrt{-g}\left( R-2\Lambda \right) +\frac{%
1}{4\kappa ^{2}\mu _{G}}\int d^{3}x\varepsilon ^{\alpha \beta \gamma
}(\Gamma _{\alpha \sigma }^{\rho }\partial _{\beta }\Gamma _{\gamma \rho
}^{\sigma }+\frac{2}{3}\Gamma _{\alpha \sigma }^{\rho }\Gamma _{\beta \eta
}^{\sigma }\Gamma _{\gamma \rho }^{\eta })+\frac{1}{2\kappa ^{2}m^{2}}\int
d^{3}x\sqrt{-g}(R_{\mu \nu }R^{\mu \nu }-\frac{3}{8}R^{2}).
\end{equation*}%
The field equations are 
\begin{equation}
R_{\mu \nu }-\frac{1}{2}Rg_{\mu \nu }-2\Lambda g_{\mu \nu }+\frac{1}{2m^{2}}%
K_{\mu \nu }+\frac{1}{\mu _{G}}C_{\mu \nu }=0,  \label{fieldeq}
\end{equation}%
where the Cotton tensor $C_{\mu \nu }$ is given by (\ref{Cotono}), while the
tensor $K_{{\mu \nu }}$ is given by 
\begin{equation}
K_{\mu \nu }=2\square {R}_{\mu \nu }-\frac{1}{2}\nabla _{\mu }\nabla _{\nu }{%
R}-\frac{1}{2}\square {R}g_{\mu \nu }+4R_{\mu \alpha \nu \beta }R^{\alpha
\beta }-\frac{3}{2}RR_{\mu \nu }-R_{\alpha \beta }R^{\alpha \beta }g_{\mu
\nu }+\frac{3}{8}R^{2}g_{\mu \nu }.  \label{eom}
\end{equation}%
The tensor $K_{\mu \nu }$ has the property that its trace equals the
Lagrangian it comes from; namely $g^{\mu \nu }K_{\mu \nu }=R_{\mu \nu
}R^{\mu \nu }-\frac{3}{8}R^{2}$. Since the Cotton tensor is traceless, we
find from (\ref{fieldeq}) that $6m^{2}\Lambda -m^{2}R-\frac{3}{8}%
R^{2}+R_{\mu \nu }R^{\mu \nu }=0$.

Because of the presence of higher-curvature terms in the action, the central
charges associated to the Virasoro algebras that generate the asymptotic
isometry group of the AdS$_{3}$ sector of the theory receive an extra term.
On the other hand, the presence of the gravitational Chern-Simons term
introduces a difference between the left-moving and the right-moving central
charges, encoding the diffeomorphism anomaly. The total central charges are
then given by%
\begin{equation}
c_{L}=\frac{3l}{2G}\left( 1-\frac{1}{\mu _{G}l}+\frac{1}{2m^{2}l^{2}}\right)
,\qquad c_{R}=\frac{3l}{2G}\left( 1+\frac{1}{\mu _{G}l}+\frac{1}{2m^{2}l^{2}}%
\right)  \label{C}
\end{equation}%
with 
\begin{equation}
\Lambda =-\frac{1}{l^{2}}\left( 1-\frac{1}{4m^{2}l^{2}}\right) .  \label{O2}
\end{equation}

The asymptotic AdS$_{3}$ boundary conditions our solution (\ref{uncharged})
fulfills are given by the following next-to-leading behavior of the metric $%
g_{\mu \nu }$,%
\begin{eqnarray}
g_{tt} &\simeq &-\frac{r^{2}}{l^{2}}+\mathcal{O}(\log (r)),\qquad
g_{rr}\simeq \frac{l^{2}}{r^{2}}+\mathcal{O}(r^{-4}),\qquad g_{\phi t}\simeq 
\mathcal{O}(\log (r)),  \label{AC} \\
g_{\phi r} &\simeq &\mathcal{O}(1),\qquad g_{\phi \phi }\simeq r^{2}+%
\mathcal{O}(\log (r)),\qquad g_{rt}\simeq \mathcal{O}(1),  \label{AC2}
\end{eqnarray}%
where the leading term corresponds to the metric of the massless BTZ
solution. We emphasize here that this notion of \textit{asymptotically AdS}$%
_{3}$\textit{\ space} is slightly different from the one originally
introduced by Brown and Henneaux for the case of three-dimensional General
Relativity. The presence of contributions of order $\mathcal{O}(\log (r))$
makes the asymptotic fall-off weaker than the one considered in \cite{BH}.
Nevertheless, these boundary conditions are still consistent with the
definition of asymoptotic charges that realize the boundary two-dimensional
conformal algebra with central charges (\ref{C}).

As mentioned above, the claim is that the uncharged solution (\ref{uncharged}%
), which incarnates asymptotic conditions (\ref{AC})-(\ref{AC2}), is also a
solution of the full theory (\ref{fieldeq}) at a special point of the space
of parameters. The metric of the solution found in Section 2 reads 
\begin{equation}
ds^{2}=-N^{2}(r)dt^{2}+\frac{dr^{2}}{N^{2}(r)}+r^{2}(N_{\phi }(r)dt-d\phi
)^{2}+N_{k}^{2}(r)(dt-ld\phi )^{2}  \label{unchargedd}
\end{equation}%
where%
\begin{equation}
N^{2}(r)=\frac{r^{2}}{l^{2}}-M+\frac{M^{2}l^{2}}{4r^{2}},\qquad \qquad
N_{\phi }(r)=\frac{Ml}{2r^{2}},
\end{equation}%
and%
\begin{equation}
N_{k}^{2}(r)=k\log |r^{2}-Ml^{2}/2|.  \label{Nkk}
\end{equation}

Notice that, unlike the way we presented the solution in Section 2, here we
have written (\ref{Nkk}) in such a way that it includes the absolute value $%
|r^{2}-Ml^{2}/2|$ in the argument of the logarithm. This makes the metric to
be well defined also in the region $r^{2}<Ml^{2}/2$. It is not hard to
verify that metric (\ref{unchargedd})-(\ref{Nkk}) solves the equations of
motion (\ref{fieldeq}) if the coupling constants satisfy the relation%
\begin{equation}
\frac{1}{\mu _{G}l}-1=\frac{1}{2m^{2}l^{2}},  \label{O}
\end{equation}%
which, as we observe from (\ref{C}), coincides with the point of the moduli
space at which the central charge associated to the left-moving Virasoro
algebra vanishes, that is $c_{L}=0$. This coincides with the result of
Section 2 in the limit $1/$ $m^{2}\rightarrow 0$, for which (\ref{O2}) and (%
\ref{O}) correspond to the chiral point $\mu _{G}=1/l$ of \cite{LSS}. The
limit $1/\mu _{G}\rightarrow 0$ of this solution was recently studied in 
\cite{Clement}.

In conclusion, we have shown that the solution to Log Gravity (\ref%
{uncharged}) actually persists if the Lagrangian of New Massive Gravity is
included in the gravitational action \footnote{%
Other solutions that persist when the NMG Lagrangian is added to TMG were
studied recently in \cite{Kundt,WarpedZZ}}. This solution incarnates the
weakened AdS$_{3}$ boundary conditions whose consistency was studied in Ref. 
\cite{Sun}. It is worth pointing out that the solution (\ref{unchargedd})-(%
\ref{Nkk}) is also persistent for a much more general deformation of the
gravity Lagrangian. Again, this persistence of the solution is easily
explained once one is reminded of the fact that it is locally equivalent to
a $pp$-wave solution, being of constant scalar curvature $R=-6/l^{2}$, $%
R_{\mu \nu }R^{\mu \nu }=12/l^{4}$. Nevertheless, the presence of a term
like $\square {R}_{\mu \nu }$ in (\ref{eom}) makes the story non-trivial.
Solutions to three-dimensional generalized massive gravity that are
conformally equivalent to $pp$-waves were recently explored in \cite%
{AdSwaves}.

\section{Acknowledgements}

The work of A.G. was supported by Universidad de Buenos Aires, UBA. The work
of G.G. is supported by UBA, CONICET and ANPCyT through grants UBACyT X861,
PIP6160, PICT34557. The work of Y.V. was partially supported by Ministerio
de Educaci\'{o}n through MECESUP Grants FSM 0204, by Direcci\'{o}n de
Estudios Avanzados PUCV and by CONICyT Scholarship 2008. Discussions on
related subjects with G. Barnich, F. Canfora, G. Comp\`{e}re, S. Detournay,
M. Henneaux, J. Oliva, R. Troncoso, and J. Zanelli are acknowledged. G.G. is
indebted to G. Cl\'{e}ment, D. Grumiller, M. Kleban and M. Porrati for
useful comments and enjoyable discussions on related subjects. He also
thanks the referee of Phys. Rev. D for interesting comments, and thanks C.
Garraffo, A. Lawrence, and the Theory Group of Martin Fisher School of
Physics, Brandeis University, for the hospitality during his stay. A.G. and
G.G. specially thank M. Leston for collaboration.


\begin{thebibliography}{99}
\bibitem{witten} E. Witten, \textit{Three-dimensional gravity reconsidered},
[arXiv:0706.3359]

\bibitem{witten2} A. Maloney and E. Witten, \textit{Quantum gravity
partition function in three-dimensions}, [arXiv:0712.0155].

\bibitem{gaberdiel} M. Gaberdiel, \textit{Constraints on extremal self-dual
CFTs}, JHEP 0711 (2007) 087, [arXiv:0707.4073].

\bibitem{gaiotto} D. Gaiotto, \textit{Monster symmetry and extremal CFTs},
[arXiv:0801.0988].

\bibitem{gaberdiel2} M. Gaberdiel and. C. Keller, \textit{Modular
differential equations and null vectors}, [arXiv:0804.0489].

\bibitem{DJT} S. Deser, R. Jackiw and S. Templeton, \textit{%
Three-Dimensional Massive Gauge Theories}, Phys. Rev. Lett. \textbf{48}
(1982) 975.

\bibitem{DJT2} S. Deser, R. Jackiw and S. Templeton, \textit{Topologically
massive gauge theories}, Annals Phys. \textbf{140} (1982) 372 [Erratum-ibid 
\textbf{185}, 406.1966 APNYA, 281, 409 (1988 APNYA, 409-449.2000)].

\bibitem{MCL} K. Ait Moussa, G. Cl\'{e}ment and C. Leygnac, \textit{The
black holes of topologically massive gravity}, Class. Quant. Grav. \textbf{20%
} (2003) L277, [arXiv:gr-qc/0303042].

\bibitem{KL} P. Kraus and F. Larsen, \textit{Holographic Gravitational
Anomalies}, JHEP \textbf{01} (2006) 022, [arXiv:hep-th/0508218].

\bibitem{LSS2} W. Li, W. Song and A. Strominger, \textit{Chiral Gravity in
Three Dimensions}, JHEP \textbf{04} (2008) 082, [arXiv:0801.4566].

\bibitem{CDWW} S. Carlip, S. Deser, A. Waldron and D. Wise, \textit{%
Cosmological Topologically Massive Gravitons and Photons}, [arXiv:0803.3998].

\bibitem{GJ} D. Grumiller and N. Johansson, \textit{Instability in
cosmological topologically massive gravity at the chiral point},
[arXiv:0805.2610].

\bibitem{S} A. Strominger, \textit{A Simple Proof of the Chiral Gravity
Conjecture}, [arXiv:0808.0506].

\bibitem{Hotta} K. Hotta, Y. Hyakutake, T. Kubota and H. Tanida,\textit{\
Brown-Henneaux's canonical approach to Topologically Massive Gravity}, JHEP 
\textbf{07 }(2008) 066, [arXiv:0805.2005].

\bibitem{LSS} W. Li, W. Song and A. Strominger, \textit{Comment on
"Cosmological Topological Massive Gravitons and Photons"}, [arXiv:0805.3101].

\bibitem{SS} I. Sachs and S. Solodukhin, \textit{Quasi-Normal Modes in
Topologically Massive Gravity}, JHEP \textbf{08} (2008) 003,
[arXiv:0806.1788].

\bibitem{C} S. Carlip, \textit{The Constraint Algebra of Topologically
Massive AdS Gravity}, [arXiv:0807.4152].

\bibitem{GKP} G. Giribet, M. Kleban and M. Porrati, \textit{Topologically
Massive Gravity at the Chiral Point is Not Unitary}, JHEP \textbf{10} (2008)
045, [arXiv:0807.4703].

\bibitem{CDWW2} S. Carlip, S. Deser, A. Waldron and D. Wise, \textit{%
Topologically Massive AdS Gravity}, Phys. Lett. \textbf{B666} (2008) 272,
[arXiv:0807.0486].

\bibitem{GJJ} D. Grumiller, R. Jackiw and N. Johansson, \textit{Canonical
analysis of cosmological topologically massive gravity at the chiral poin}t,
Contribution to Wolfgang Kummer Memorial Volume, [arXiv:0806.4185].

\bibitem{GJ2} D. Grumiller and N. Johansson, \textit{Consistent boundary
conditions for cosmological topologically massive gravity at the chiral point%
}, [arXiv:0808.2575].

\bibitem{Park} Mu-in Park, \textit{Constraint Dynamics and Gravitons in
Three Dimensions}, JHEP \textbf{09} (2008) 0842008, [arXiv:0805.4328].

\bibitem{Desernuevo} S. Deser, \textit{Distended Topologically Massive
Electrodynamics}, Contribution to Wolfgang Kummer Memorial Volume,
[arXiv:0810.5384].

\bibitem{GHSS} M. Guica, T. Hartman, W. Song and A. Strominger, \textit{The
Kerr/CFT Correspondence}, [ arXiv:0809.4266].

\bibitem{Suecos} I. Bengtsson and P. Sandinar, \textit{Anti-de Sitter space,
squashed and stretched}, Class. Quant. Grav. \textbf{23} (2006) 971,
[aXiv:gr-qc/0509076].

\bibitem{ALPSS} D. Anninos, W. Li, M. Padi, W. Song and A. Strominger, 
\textit{Warped AdS}$_{\mathit{3}}$\textit{\ Black Holes}, [arXiv:0807.3040].

\bibitem{CD} G. Comp\`{e}re and S. Detournay, \textit{Semi-classical central
charge in topologically massive gravity}, [arXiv:0808.1911].

\bibitem{Pope} H. Lu, Jianwei Mei and C. Pope, \textit{Kerr/CFT
Correspondence in Diverse Dimensions}, [arXiv:0811.2225].

\bibitem{Stromingernuevo} T. Hartman, K. Murata, T. Nishioka and A.
Strominger, \textit{CFT Duals for Extreme Black Holes}, [arXiv:0811.4393].

\bibitem{eloy} E. Ay\'{o}n-Beato and M. Hassa\"{\i}ne, \textit{pp Waves of
Conformal Gravity with Self-Interacting Source}, Annals Phys. \textbf{317}
(2005) 175, [arXiv:hep-th/0409150].

\bibitem{BTZ} M. Ba\~{n}ados, C. Teitelboim and J. Zanelli,\textit{The black
hole in three-dimensional spacetime}, Phys. Rev. Lett. \textbf{69} (1992)
1849, [arXiv:hep-th/9204099].

\bibitem{BTZ2} M. Ba\~{n}ados, M. Henneaux, C. Teitelboim and J. Zanelli, 
\textit{Geometry of the 2+1 Black Hole}, Phys. Rev. \textbf{D48} (1993)
1506, [arXiv:gr-qc/9302012].

\bibitem{Clementcomment} The authors thank G\'{e}rard Cl\'{e}ment for useful
discussions about this point.

\bibitem{BH} J. Brown and M. Henneaux, \textit{Central Charges in the
Canonical Realization of Asymptotic Symmetries : An example from
Three-Dimensional Gravity}, Commun. Math. Phys. \textbf{104} (1986) 207.

\bibitem{footnote} It is worth pointing out that this logarithmic
next-to-leading order is not the one considered in \cite{HMTZ, HMTZ2}; c.f.
the leading behaviour of $g_{\phi t}$.

\bibitem{HMTZ} M. Henneaux, C. Mart\'{\i}nez, R. Troncoso and J. Zanelli, 
\textit{Asymptotically Anti-de Sitter spacetimes and scalar fields with a
logarithmic branch}, Phys. Rev. \textbf{D70} (2004) 044034,
[arXiv:hep-th/0404236].

\bibitem{HMTZ2} M. Henneaux, C. Mart\'{\i}nez, R. Troncoso and J. Zanelli, 
\textit{Black holes and asymptotics of 2+1 gravity coupled to a scalar field}%
, Phys.Rev . \textbf{D65} (2002) 104007, [arXiv:hep-th/0201170].

\bibitem{BK} V. Balasubramanian and P. Kraus, \textit{A Stress Tensor for
Anti-de Sitter Gravity}, Commun. Math. Phys. \textbf{208} (1999) 413,
[arXiv:hep-th/9902121].

\bibitem{BY} J. Brown and J. York, \textit{Quasilocal energy and conserved
charges derived from the gravitational action}, Phys. Rev. \textbf{D47}
(1993) 1407.

\bibitem{AGM} A. Garbarz, G. Giribet and M. Leston, in preparation.

\bibitem{MZ} C. Mart\'{\i}nez and J. Zanelli, \textit{Conformally dressed
black hole in 2+1 dimensions}, Phys. Rev. \textbf{D54} (1996) 3830,
[arXiv:gr-qc/9604021].

\bibitem{CM} K. Ait Moussa and G. Cl\'{e}ment, \textit{Topologically massive
gravito-electrodynamics: exact solutions}, Class. Quant. Grav. \textbf{13}
(1996) 2319, [arXiv:gr-qc/9602034].

\bibitem{unidades} The reader may wonder about the units of $l^{2}$ in Eq. (%
\ref{uv}). In this section we are using the conventions of \cite{CM} with
the intention to make the comparison easier.

\bibitem{C2} G. Cl\'{e}ment, \textit{Particle-like solutions to
topologically massive gravity}, Class. Quant. Grav. \textbf{11} (1994) L115,
[arXiv:gr-qc/9404004].

\bibitem{BC} A. Bouchareb and G. Cl\'{e}ment, \textit{Black hole mass and
angular momentum in topologically massive gravity}, Class. Quant. Grav. 
\textbf{24} (2007) 5581, [arXiv:0706.0263].

\bibitem{Clementcomment2} More precisely, the agreement is up to an
irrelevant numerical factor. The authors are grateful to G. Gl\'{e}ment for
discussing the details of this computation.

\bibitem{Clementnuevo} K. Ait Moussa, G. Cl\'{e}ment, H. Guennoune and C.
Leygnac, \textit{Three-dimensional Chern-Simons black holes}, Phys. Rev. 
\textbf{D78} (2008) 064065, [arXiv:0807.4241].

\bibitem{Troncosocomment} The authors thank Ricardo Troncoso for pointing
out the relation with the $pp$-wave solution.

\bibitem{Gibbons} G. Gibbons, C. Pope and E. Sezgin, \textit{The General
Supersymmetric Solution of Topologically Massive Supergravity}, Class.
Quant. Grav. \textbf{25} (2008) 205005, [arXiv:0807.2613].

\bibitem{NMG} E. Bergshoeff, O. Hohm and P. Townsend, \textit{Massive
Gravity in Three Dimensions}, Phys. Rev. Lett. \textbf{102} (2009) 201301,
[arXiv:0901.1766].

\bibitem{Sun} Y. Liu and Y-W. Sun, \textit{On the Generalized Massive
Gravity in $AdS_3$}, Phys. Rev. \textbf{D79} (2009) 126001,
[arXiv:0904.0403].

\bibitem{Maloney} A. Maloney, W. Song and A. Strominger, \textit{Chiral
Gravity, Log Gravity and Extremal CFT}, [arXiv:0903.4573].

\bibitem{Clement} G. Cl\'{e}ment, \textit{Black holes with a null Killing
vector in three-dimensional massive gravity}, Class. Quantum Grav. \textbf{26%
} (2009) 165002, [arXiv:0905.055].

\bibitem{Kundt} M. Chakhad, \textit{Kundt spacetimes of massive gravity in
three dimensions}, [arXiv:0907.1973].

\bibitem{WarpedZZ} G. Cl\'{e}ment, \textit{Warped AdS$_{3}$ black holes in
new massive gravity}, Class. Quantum Grav. \textbf{26} (2009) 105015,
[arXiv:0902.4634].

\bibitem{AdSwaves} E. Ay\'{o}n-Beato, G. Giribet and M. Hassa\"{\i}ne, 
\textit{Bending AdS Waves with New Massive Gravity}, JHEP \textbf{0905}
(2009) 029, [arXiv:0904.0668].
\end{thebibliography}
\end{document}